    \documentclass[preprint]{aastex}

    \newcommand{\chandra}{{\it Chandra X-Ray Observatory}}
    \newcommand{\nas}{the National Aeronautics and Space Administration}
    \newcommand{\hst}{{\it Hubble Space Telescope}}
    \newcommand{\stsci}{Space Telescope Science Institute}
    \newcommand{\ergs}{ergs~s$^{-1}$}
    \newcommand{\ecs}{ergs~cm$^{-2}$~s$^{-1}$}
    \newcommand{\msun}{M$_{\odot}$}
    \newcommand{\lsun}{L$_{\odot}$}

\begin{document}
    \title{X-Ray and Infrared Observations of Embedded Young Stars in L1630  }
    \author{Theodore Simon\footnotemark[1], Sean M. Andrews\footnotemark[1], 
	 and John T. Rayner\footnotemark[1]}
    \affil{Institute for Astronomy, University of Hawaii\\ 2680 Woodlawn Drive,
       Honolulu, HI 96822}

    \author{Stephen A. Drake}
    \affil{HEASARC, NASA/GSFC, Greenbelt, MD 20771}

    \footnotetext[1]{Visiting Astronomer at the Infrared Telescope Facility, which is 
     operated by the University of Hawaii under contract from NASA.}

    \begin{abstract}
	The HH\,24-26 star forming region within the Lynds 1630 (L1630) dark cloud in Orion 
	contains a remarkable collection of rare Class~0 and Class~I protostars, collimated 
	molecular and ionized jets, and a luminous but spatially unresolved ASCA X-ray source. 
	To study the X-ray properties of the embedded protostar population of that region, 
	we  have obtained a deep X-ray image with the ACIS-S camera on board the {\it Chandra 
	X-Ray Observatory}.  A number of H$\alpha$ emission-line objects were detected in the
	areas surrounding HH\,24--26, of which the weak-line T~Tauri star SSV\,61 was the 
	brightest source, at a steady luminosity of $L_x$(0.3--10 keV)$ = 10^{31.9}$ \ergs. 
	Two Class~I protostars aligned with optical jets in HH\,24, SSV\,63E and SSV\,63W, 
	were also detected, as was the continuum radio source SSV\,63NE, which is very likely 
	an extreme Class~I or Class~0 object. We observed no X rays from the Class~0 protostars 
	HH\,24-MMS and HH\,25-MMS, nor any from regions of the cloud bounded by HH\,25 and 
	HH\,26, at a $2\sigma$ upper limit of $L_x \sim 10^{30.0}$ \ergs. HH\,26-IR, the 
	Class~I object thought to be the origin of the HH\,26 flow, was not detected.  
	Near-infrared spectroscopy obtained at the NASA IRTF reveals 3\,$\micron$ ice bands 
	in the spectra of SSV\,59, 63E, 63W, and HH\,26-IR, and 2.3\,$\micron$ CO overtone 
	absorption bands for SSV\,61. SSV\,60, which lies astride one end of the great arc 
	of nebulosity forming HH\,25, exhibits a deep infrared ice band and CO absorption, 
	but is not an X-ray source, and is most likely a distant background giant of late 
	spectral type.             

    \end{abstract}


    \section{INTRODUCTION}

	The Lynds 1630 (L1630) dark cloud is part of the Orion B Molecular Cloud.  At a distance 
	of approximately 450 pc, it is one of the nearest giant molecular clouds.  Its mass is 
	estimated to be more than $\sim1 \times 10^5$ \msun\  (Maddalena et al.  1986). 

	L1630 shows many signs of on-going star formation: ({\it a}) dense cores of molecular gas 
	and high velocity outflows (Snell \& Edwards 1982; Gibb \& Heaton 1993; Gibb et al. 1995);
	({\it b}) bright reflection nebulae (e.g., NGC 2068 and NGC 2071); ({\it c}) T Tauri stars
	(Herbig \& Kuhi 1963; Wiramihardja et al. 1989); ({\it d}) Herbig-Haro (HH) objects, including 
	the morphologically complex HH\,24--26 (Herbig 1974; Solf 1987; Mundt, Ray, \& Raga 1991;  
	Eisl\"{o}ffel \& Mundt 1997); ({\it e}) luminous far-infrared and submillimeter sources 
	(Cohen et al. 1984; Cohen \& Schwartz 1987; Gibb \& Heaton 1993; Schwartz, Burton, 
	\& Herrmann 1997; Mitchell et al. 2001); ({\it f}) radio continuum sources (Bontemps, 
	Andr\'{e}, \& Ward-Thompson 1995; Verdes-Montenegro \& Ho 1996; Reipurth et al. 2002); 
	and ({\it g}) dense clusters of optically obscured near-infrared sources (Strom et al. 
	1975; Strom, Strom, \& Vrba 1976). The number density of young stars in L1630, 450 
	pc$^{-3}$, is comparable with that of the $\rho$~Ophiuchus star-forming region (Lada 
	et al. 1991).  

	Starting with the work of the Stroms, L1630 has been studied in ever increasing detail at 
	infrared wavelengths. Particular attention has been focused recently on its many outflows 
	and HH features.  Evidence for emission in the 2\,$\micron$ near-infrared lines of H$_2$  
	from parsec-scale HH flows and jets has been presented by Zealey et al. (1992), and 
	more extensively by Davis et al. (1997) and Chrysostomou et al. (2000). Images of HH\,24 
	in narrow emission lines at optical and infrared wavelengths trace out at least three 
	intersecting, collimated, bipolar flows (e.g., Mundt et al. 1991:  Fig. 17--19; Davis et 
	al. 1997:  Fig. 2). The flows emanate from the vicinity of HH\,24B, at the location of the 
	bright near-infrared sources SSV\,63E and SSV\,63W. Both objects are presumed from their 
	steep spectral energy distributions to be Class I protostars (Cohen et al. 1984), that is, 
	they are in a very early stage of evolution preceding the T Tauri star (TTS) or so-called 
	Class II phase, and in the process of actively accreting material from their protostellar 
	disks. The dynamical age of HH\,24 is of the order of $\sim10^4$~yr (Eisl\"{o}ffel \& Mundt 
	1997; Snell \& Edwards 1982), which implies that the Class I objects powering those flows 
	must be very young, and also in a very short-lived phase of evolution. 

	In the past, the bright HH\,24A knot was interpreted as the ``working surface'' or bow 
	shock for one of those outflows (Mundt et al. 1991; Solf 1987), but it is now recognized 
	(e.g., Eisl\"{o}ffel \& Mundt 1997) to be the terminus of an independent outflow, which
	originates in a region immediately to the southwest, at the location of a deeply embedded 
	Class~0 protostar, HH\,24-MMS. Class~0 objects like HH\,24-MMS represent the main accretion 
	phase for protostars and are extremely rare because their lifetimes are so short. HH\,24-MMS 
	is observed directly at near-infrared wavelengths only in shocked H$_2$ (Bontemps, 
	Ward-Thompson, \& Andr\'{e} 1996; Davis et al. 2002). It was discovered as a bright source 
	at submillimeter and radio wavelengths 
	(Chini et al. 1993; Ward-Thompson et al. 1995; Bontemps et al. 1995). Greybody fits suggest 
	a color temperature of 20~K and a total luminosity of $\sim$5--20 \lsun, of which $\sim2.5$ 
	\lsun\ originates from an unresolved and possibly disk-like central source (Ward-Thompson et 
	al. 1995). The compact source is surrounded by a much more extended envelope, from which it
	is most likely accreting material at a very high rate (Chandler et al. 1995b). The detection 
	of an H$_2$ jet, blue-shifted CO emission (Bontemps et al. 1996), and indications of disk 
	structure suggest that HH\,24-MMS, even at its remarkably young age, has already begun to 
	drive a powerful collimated outflow into the surrounding molecular cloud.  Various theories
	and mechanisms have been advanced in the past to explain such highly focused (and frequently
	bipolar) jets. In the class of magnetic models, for example, it has been proposed that the  
	outflow is confined and driven by interactions between the magnetosphere of the protostar 
	and the primordial field that threads its accretion disk (e.g., Shu et al. 1994).   
                
	South of HH\,24, high resolution radio, millimeter, and submillimeter maps have revealed a
	number of strong molecular peaks and bipolar outflows in the proximity of HH\,25 and HH\,26 
	(Gibb \& Heaton 1993; Gibb et al. 1995; Gibb \& Davis 1998).  The surrounding areas have been 
	extensively mapped in the $v = 1-0\ S$(1) line of H$_2$ at 2\,$\micron$ (Davis et al. 1997;
	Chrysostomou et al. 2000; Davis et al. 2002). A Class~0 object, HH\,25-MMS, was discovered 
	$10\arcsec$ south of the optically-visible knot HH\,25A as both a 1.3-mm and a 3.6-cm source 
	(Bontemps et al. 1995). Strong 450 and 850\,$\micron$ continuum emission has been observed 
	from the same location (Mitchell et al. 2001; Phillips, Gibb, \& Little 2000). The brightest 
	near-infrared objects in the vicinity of HH\,25--26 are SSV\,59, SSV\,60, and HH\,26-IR (Strom 
	et al. 1976; Davis et al. 1997). None of the three has been shown conclusively to be the origin 
	of a large-scale flow, although the weight of the evidence for HH\,26 (in particular, the 
	discovery of a compact H$_2$ jet:  Davis et al. 2002) now points most convincingly to HH\,26-IR. 
	The latter, as well as SSV\,59, is generally regarded as a Class~I protostar (e.g., Bontemps et 
	al. 1995, Davis et al. 2002). The status of SSV\,60 is unknown; unlike the other two sources, it 
	is associated with neither a submillimeter peak (Lis, Menten, \& Zylka 1999; Mitchell et al. 2001; 
	Phillips et al. 2001) nor a molecular core (e.g., Torrelles et al. 1989; Gibb \& Heaton 1993). 
	
	The same process that drives the outflows and jets of Young Stellar Objects (YSOs) is expected 
	to result in the repeated shearing and reconnection of  the magnetic field lines that attach 
	such a star to its circumstellar disk, and thereby to generate a substantial amount of X-ray 
	emission  (Hayashi, Shibata, \& Matsumoto 1996; Shu et al. 1997; Goodson, Winglee, \& B\"{o}hm 
	1997). The numerical simulations of Hayashi et al. and Goodson et al. in particular demonstrated 
	that as a consequence of reconnection events some of the magnetically entrapped gas that is 
	close to an accreting YSO (not necessarily its outflowing jet) will reach temperatures as high 
	as 100 MK and will be flung outwards, away from the star. Such high temperatures should give
	rise to hard X-ray spectra. X-ray emission at equally high temperatures can also be produced 
	in an accretion shock at the footpoints of the magnetic field, where infalling material 
	from the disk streams down onto the surface of the forming star (Uchida \& Shibata 1984). 
%
%
	The HH\,24--26 region was observed in X-rays with ASCA for a net exposure time of 30 
	kiloseconds (ks) by Ozawa et al. (1999). The strongest source in the soft X-ray band 
	(0.7--2.0 keV) in that observation was SSV\,61, a T~Tauri star (TTS) which is embedded in 
	an unusual patch of nebulosity situated mid-way on the sky between HH\,24 and HH\,25--26 
	(e.g., Scarott, Gledhill, \& Warren-Smith 1987). Strom et al. (1975) classified the star as 
	an M2--M4 giant from image-tube spectra. A CCD spectrum of SSV\,61,  kindly made available 
	to us by G. Herbig, exhibits H$\alpha$ emission and \ion{Li}{1} $\lambda6707$ absorption 
	equivalent widths that are typical of the class of weak-lined T Tauri stars (wTTS), EW(H$\alpha$) 
	= $-5$ \AA\ and EW(\ion{Li}{1}) = 480 m\AA\ (cf. Walter 1986; Mart\'{\i}n et al. 1994).   
	The brightest emission in the hard ASCA band (2.0--8.0 keV) is located north of SSV\,61, 
	in the vicinity of HH\,24. The high-energy peak lies nearest SSV\,63, but cannot be 
	identified unambiguously with either of its infrared components due to the low spatial 
	resolution of ASCA. Ozawa et al. noted the great difficulty of separating the X-ray flux 
	of SSV\,63 from that of SSV\,61, given their offset of just $1\farcm8$, at roughly the 
	half-power radius of ASCA.  The quiescent luminosity $L_x \simeq 10^{32}$ \ergs\ they 
	derive for SSV\,63 places it above the top end of the range thus far observed for all but 
	a few Class I objects, by factors of 50--100 (Carkner, Kozak, \& Feigelson 1998; Preibisch, 
	Neuh\"{a}user, \& Stanke 1998). Whether SSV\,63 is truly one of the most X-ray luminous 
	YSOs in the sky, however, has awaited confirmation from high resolution imaging that resolves 
	not only SSV\,61 from SSV\,63, but also the individual infrared components of SSV\,63 itself. 
	That task, including the requisite high precision astrometry, has now been accomplished by 
	the {\it Chandra} observations reported here. 

	As for the Class~0 sources in L1630, Ozawa et al. found no sign of X-ray emission at the 
	position of HH\,24-MMS, located in the saddle between the soft and hard X-ray peaks in 
	their image, nor any emission at the position of HH\,25-MMS, which was also in the ASCA 
	field of view. Their $3\sigma$ upper limit of $L_x \lesssim 1.2\times 10^{31}$ \ergs\ is 
	not a stringent limit on the emission, however, since YSOs elsewhere in Orion and in other 
	star forming regions have been detected at much fainter levels, $L_x \simeq \mbox{few} 
	\times 10^{30}$ \ergs\ (e.g., Feigelson et al. 2002; Imanishi, Koyama, \& Tsuboi 2001).
	Using the {\it Chandra Observatory}, we have undertaken a far more sensitive 
	search for the onset of X-ray emission in the Class 0 objects within the L1630 cloud.  
	Our objective was to establish an evolutionary progression in X-ray properties within 
	a single cloud, and to examine the possible relationship between protostellar X rays, 
	accretion disks, magnetically confined jets, and large-scale molecular outflows.  Taking 
	advantage of the lower background, greater sensitivity, and unsurpassed spatial resolution
	of {\it Chandra}, we are now able to set a 10 times smaller bound on possible X-ray 
	emission from the three established Class 0 protostars in the L1630 cloud, HH\,24-MMS, 
	HH\,25-MMS, and LBS\,17-H.

    \section{OBSERVATIONS and RESULTS}

    \subsection{ The X-Ray Imaging-Spectroscopy }

	L1630 was observed with the Advanced CCD Imaging Spectrometer array (ACIS: Garmire et 
	al. 2002) on the {\it Chandra X-Ray Observatory} for a net exposure time (livetime) of 
	62.8 ks.  The date of observation was 14--15 November  2002. All six detector chips of 
	the ACIS-S camera, S0--S5, were activated during the observation. The aim point of the 
	observation was chosen to place SSV\,61 close to the center of the back-illuminated S3 
	chip and to put HH\,24 and HH\,25--26 on the same device within a few arcminutes of the 
	optical axis of the telescope, thereby ensuring the best optical performance for our 
	highest priority targets within the ACIS field of view. The roll angle of the observation, 
	$61^\circ$, was selected to position the NGC 2068 reflection nebula, located to the 
	northeast of HH\,24--26, on the back-illuminated S1 chip of the detector array and the 
	Class 0 object, NGC 2068/LBS\,17-H (Gibb \& Little 2000), on the adjacent, front-illuminated 
	S2 chip. The majority of the field of view is shown in Fig. 1, superposed on the red image 
	of the Digitized Sky Survey of the \stsci\ (DSS). The approximate locations of  HH objects
	in the neighboring areas are marked, as are the positions of the two dozen X-ray sources
	we have detected in our {\it Chandra} observation.         
	
	To analyze the {\it Chandra} image, we initially filtered the Level-2 X-ray event file 
	provided by the pipeline data processing system, retaining only the photon events that were 
	assigned ASCA grades 0, 2, 3, 4, and 6 and that were within the broad (0.5--10) keV energy 
	band in which the ACIS detector has its highest response and the background radiation is at 
	its lowest levels.  Subsequent steps in the analysis, including photometry and astrometry, were 
	carried out following various data analysis threads within CIAO, the interactive {\it Chandra} 
	data analysis package.  For source detection, we applied the {\sc celldetect} sliding-cell 
	detection algorithm. Twenty-five sources, all of them point-like, were found on the six chips.
	To extract X-ray counts for each source, we used the {\sc dmextract} tool. We employed 
	a circular extraction region with a radius of 10 pixels ($r = 5\arcsec$) to measure the 
	brightness of sources located within 7\arcmin\ of the optical axis of the telescope, a radius 
	twice as large for sources more than 7\arcmin\ off-axis, and an annulus at least 10 times as 
	large to evaluate the background emission.

	The results of the photometry are given in Table 1. Listed there are the right ascension 
	and declination (J2000) of each source, its off-axis angle, the net source counts extracted
	within the 0.5--10 keV band, and the statistical errors in the source counts. In the final 
	column we tabulate the count rate for the effective exposure time of each source, which 
	takes account of the variations in the sensitivity of the camera across the field of view, 
	as indicated by our exposure map calculations. The adjustments are negligible for the 
	nearly on-axis sources on the S3 chip that are the main interest of the present study, but 
	may be as great as 50\% for the few objects that are located either near the edge of a 
	detector chip or else far away from the optical axis of the telescope. We also list possible 
	optical and infrared counterparts, which were found by cross-matching the X-ray coordinates 
	with entries in the {\sc simbad} database and by visual comparison with red images from 
	the DSS as well as infrared images from the 2MASS sky survey. The identifications include 
	several H$\alpha$ emitters (Herbig \& Kuhi 1963), named infrared sources (Strom et al. 1976), 
	anonymous near-infrared objects from the 2MASS survey, and entries in the \hst\ Guide Star 
	Catalog-II (identified here with the prefix of S). The correspondence of each X-ray and 
	optical/infrared position is within 1\arcsec\ (= 2 ACIS pixels) or less. The position 
	of X-ray source no. 4 coincides with the location of a previously known non-thermal, 
	extragalactic radio source (Bontemps et al. 1995; Anglada et al. 1998). A search of the 
	NASA/IPAC Extragalactic Database (NED) turned up no other background objects in the ACIS 
	field of view. 

	SSV\,63 is resolved into three separate X-ray components, which correspond to the VLA 
	3.6-cm radio positions of SSV\,63E, 63W, and 63NE (Reipurth et al. 2002). An overlay of the 
	radio source positions on the 2--8 keV hard-band X-ray image from {\it Chandra} is shown 
	in Fig. 2. Within the uncertainties in the absolute coordinates of the radio and infrared 
	sources (see Table 1 of Bontemps et al. 1995), and the astrometric error in {\it Chandra} 
	observations (nominally 1 pixel, or $0\farcs5$, for positions close to the optical axis of 
	the telescope\footnote[2]{http://cxc.harvard.edu/cal/ASPECT/celmon/}), the X-ray centroids 
	coincide with the infrared counterpart of SSV\,63E and the brighter, southern, infrared 
	component of SSV\,63W. The fainter infrared component of SSV\,63W, located 2\arcsec\ farther 
	north (see Fig. 2 of Davis et al. 2002), appears dark in X rays.  SSV\,63NE is a source of 
	both radio emission and X-ray emission, but has no near-infrared counterpart (as we will 
	show below in \S2.2). If not a distant background object, it may qualify on that basis 
	as an extreme Class I or possible Class~0 YSO. All three components of SSV\,63 have hard 
	X-ray spectral indices:  96\% of the X-ray photons detected from SSV\,63W have pulse-height 
	energies greater than 2 keV, while all of those detected for SSV\,63E and NE have energies 
	of 2 keV or more.  We performed individual $\chi^2$ timing analyses for the three X-ray components, 
	using the {\sc xronos} software package from the High Energy Astrophysics Science Archive 
	Research Center (HEASARC) of the NASA/Goddard Space Flight Center.  We found no indication 
	of variability on timescales of 6 ks or longer, certainly nothing comparable in amplitude or 
	duration to the order-of-magnitude outburst in X-ray flux recorded by ASCA (Ozawa et al. 1999). 
	The X-ray count rates listed in Table 1 therefore likely correspond to the quiescent activity 
	levels of those sources. Changes on timescales shorter than 6 ks were not investigated owing 
	to the faintness of the sources.
	
	The strongest X-ray source in the {\it Chandra} field of view was the wTTS, SSV\,61, for 
	which we accumulated very nearly 13,000 counts in the 0.5--10 keV bandpass. The X-ray light 
	curve of SSV\,61 was stable throughout the entire observation; various $\chi^2$ tests performed 
	with {\sc xronos} found no evidence of variability on timescales greater than 300 s. To analyze 
	the X-ray spectrum of SSV\,61, we used the HEASARC {\sc xspec} spectral analysis package
	in order to fit simple multi-temperature thermal plasma models to the pulse height distribution 
	of the extracted photons. The quality of the fit was determined by $\chi^2$ minimization. In 
	the calculations, we removed a small ($<1\%$) contribution of emission from the background 
	and also updated the instrumental response function to compensate for the on-orbit decline 
	in the low-energy sensitivity of the ACIS 
	detectors\footnote[3]{http://cxc.harvard.edu/cal/Acis/Cal\_prods/qeDeg/index.html}. 

	In terms of the reduced $\chi^2$ value of the fit, no satisfactory model was found. Typical 
	of the results we obtained are the following parameters for a two-temperature, reduced-metal 
	MEKAL fit  (Mewe, Kaastra, \& Liedahl 1995), which we matched to the observed counts in the 
	0.5--7.0 keV pulse-height energy channels:  reduced $\chi^2_{(\nu=254)} = 1.45$;  $kT_1 = 
	1.23\pm0.06$ keV, $kT_2 = 10\pm3$ keV; one-third solar elemental abundance, Z = $0.33\pm0.10$; 
	hydrogen column density, N$_H = 0.30\pm0.02 \times 10^{22}$ cm$^{-2}$; volume emission
	measure, $\log EM = 54.66\pm0.05$, three-quarters of which are due to the high-temperature 
	component; broadband flux at Earth, $f_x$(0.3--10 keV) = $2.25\pm0.09
	\times 10^{-12}$ \ecs; and intrinsic (absorption-corrected) 
	X-ray luminosity for an assumed distance of 450 pc, $L_x$(0.3--10 keV) = $7.20 \times 10^{31}$ 
	\ergs.  Here, $\nu$(=254) is the number of degrees of freedom in the model. Closely similar 
	fits were obtained with both the {\sc xspec} PILEUP model (see Davis 2001), which we used 
	to account for the $\sim10$\% pileup fraction in our spectrum of SSV\,61, and also the APEC 
	plasma emission code described by Smith et al. (2001).  An example of an APEC model fit 
	is shown in Fig. 3. Except for a few possible camera artifacts, the largest residuals in
	all the model fits are attributable to excess emission in the observed spectrum near photon 
	energies of 1 keV.  The apparent excess may be due to lines of Ne IX and X or to inner
	shell transitions of high ionization stages of Fe. A similar excess has also been observed 
	in the ACIS spectra of pre-main sequence stars in other star-forming regions, e.g., Orion 
	and $\rho$ Ophiuchus (Feigelson et al. 2002; Imanishi, Tsujimoto, \& Koyama 2002).  

	An evaluation of the intrinsic 0.3--10 keV brightness of SSV\,61 for a  variety of {\sc xspec} 
	models produced very nearly the same results for the X-ray flux and for the luminosity. The 
	various  models also yielded closely similar amounts for the hydrogen column density. From 
	the models and the empirical relation of Bohlin, Savage, \& Drake (1978), we infer a visual 
	extinction of $A_v = 1.6$ mag. Our estimate, drawn from the {\it Chandra} spectrum of SSV\,61,  
	is a factor of four less than the extinction Ozawa et al. (1999) derived from their ASCA 
	observation, but is consistent with the conclusion reached by Strom et al. (1975) on the 
	basis of infrared colors that the foreground reddening to this star is relatively small. 
	Infrared spectroscopy (discussed below) further confirms the low extinction for this star.

	The three X-ray sources identified with SSV\,63 were too faint for detailed spectral 
	model fitting,  and so we limited ourselves to a few simple models of the strongest 
	component,  SSV\,63W. For that source we chose an isothermal, one temperature (1-T) 
	MEKAL model, constrained it to a sub-solar abundance of Z = 0.5, and allowed {\sc xspec} 
	to search for values of N$_H$ and $kT$ that minimized the $\chi^2$ of the fit. The best-fit 
	values derived by {\sc xspec} were N$_H = 9.0\pm1.8 \times 10^{22}$ cm$^{-2}$ (equivalent 
	to 48 mag of visible extinction) and $kT = 1.62\pm0.41$ keV. The resulting hydrogen column 
	density is at the low end of the range of values deduced by Ozawa et al. (1999), while our 
	temperature is a factor of 3 lower than their best-fit value for a quiescent interlude in 
	the ASCA observation.  A reddened bremsstrahlung model, which we chose as an alternative 
	fit, produced a similar $\chi^2$, temperature, and N$_H$ as the MEKAL model. Both models, 
	the MEKAL and the bremsstrahlung, yielded a comparable 0.3--10 keV flux and absorption-corrected 
	luminosity for SSV\,63W: $f_x \approx 5.3\pm0.5 \times 10^{-14}$  \ecs\ and $L_x \approx 1.4 
	\times 10^{31}$ \ergs.  Raising the temperature to a fixed value of $kT = 5$ keV for consistency 
	with the ASCA result reduced the column density to $N_H = 5.2\pm0.6 \times 10^{22}$ cm$^{-2}$, 
	corresponding to an extinction of $A_v = 27$ mag, but yielded a much poorer fit to the 
	{\it Chandra} pulse-height spectrum in terms of its $\chi^2$ value.

	No model analysis was possible for the two remaining components of SSV\,63, given the 
	small numbers of counts recorded for them during our {\it Chandra} observation.  To 
	estimate X-ray luminosities for those sources, we scaled from SSV\,63W in proportion to the 
	background-adjusted number of counts detected for each source. In doing so, we implicitly 
	assumed that the optimal {\sc xspec} models of SSV\,63E and SSV\,63NE would yield the same 
	parameters as the best-fitting model for SSV\,63W.  Our normalization procedure is equivalent 
	to using a counts-to-energy conversion factor (ECF) of $2.00 \times 10^{-11}$ \ecs\ per count 
	s$^{-1}$ for the X-ray flux observed at Earth, and $5.40 \times 10^{33}$ \ergs\ per count 
	s$^{-1}$ for the extinction-corrected broadband luminosity.\footnote[4]{The ECF based on our 
	spectral fits to SSV\,61 would be $0.90 \times 10^{-11}$ \ecs\ per count s$^{-1}$ for the X-ray 
	flux and $2.85 \times 10^{32}$ \ergs\ per count s$^{-1}$ for the luminosity. These smaller 
	conversion factors may be appropriate to other lightly reddened, visible stars that are listed 
	in Table 1 along with SSV\,61 but do not apply to such heavily reddened sources as SSV\,63.} 
	Anticipating the results of \S3.2, we note that sub-millimeter observations of L1630 corroborate 
	the high extinction value provided by our X-ray analysis of SSV\,63W.  On the other hand, 
	spectroscopy of the infrared ice band in the direction of SSV\,63E and W (discussed in \S2.3) 
	suggests that the foreground extinction towards both sources may be only half as large, $A_v 
	\approx 20$ mags. The apparent discrepancy raises the possibility that the X-ray sources may be 
	more deeply obscured than are the near-infrared sources, reflecting, perhaps, an anomalous 
	gas-to-dust ratio along the line of sight. 
	 
  	The X-ray luminosities for SSV\,61 and SSV\,63 are summarized in Table 2. Our $L_x$ value for 
	SSV\,61 is 15\% higher than the ASCA result of Ozawa et al. (1999) but lies substantially above 
	the high end of the range found in an extensive survey of wTTS in Taurus by ROSAT (Neuh\"{a}user 
	1997) and is well above the range of X-ray luminosities measured by {\it Chandra} for low-mass stars 
	in both Ophiuchus (Imanishi et al. 2001) and the Orion Nebula Cluster (Feigelson et al. 2003). Our 
	$L_x$ value for SSV\,63W falls near the top end of those three luminosity distributions, while the 
	values for the two other detected Class I YSOs, SSV\,63E and 63NE, are in the top quartile of each 
	range. Also listed in the table are upper limits on the X-ray emission of the near-infrared objects 
	SSV\,60 and 59, the former lying between HH\,25C and 25D along the curving ribbon of emission knots 
	that define HH\,25 (being closer to D), the latter lying just north of HH\,26A. The upper limits 
	on the count rates given in the table are three times the Poisson uncertainties in the total counts 
	(uncorrected for background emission) measured through a 10\arcsec\ diameter circular detection cell, 
	which we centered on the respective 2MASS positions listed in the second and third columns of the 
	table. Based on Eqn. 10 in Gehrels (1986), the significance level of these upper limits corresponds 
	to $2\sigma$, i.e., to a confidence level of 0.95. The upper limits obtained for a smaller 5\arcsec\ 
	diameter detection cell, which still encircles more than 95\% of the energy in the {\it Chandra} point 
	spread function,  are $\sim0.23$  dex lower at a confidence level of 0.95, in accordance with either 
	Gehrels (1986) or the Bayesian approach of Kraft, Burrows, \& Nousek (1991), but may be subject to a 
	somewhat greater uncertainty due to the imprecision of the infrared coordinates of both objects. For 
	lack of a more compelling choice, the conversion from count rate to luminosity in Table 2 is based 
	on the ECF derived from our {\sc xspec} modelling of SSV\,63W.  The upper limits cited on the 
	luminosity assume both objects lie at a distance of 450 pc within the L1630 cloud, which may not 
	be true for SSV\,60, as we discuss further in \S2.3.

	In similar fashion, we derived $2\sigma$ upper limits on the possible X-ray emission from the three
	Class 0 objects in the L1630 cloud, HH\,24-MMS, HH\,25-MMS, and LBS 17-H. Those results are also 
	given in Table 2. In each case, the detection cell was centered at the known long-wavelength or 
	radio position of the object. The J2000 coordinates of HH\,24-MMS and HH\,25-MMS were precessed 
	from the B1950 VLA positions of Bontemps et al. (1995) and Gibb (1999); those of LBS 17-H were 
	precessed from the B1950 location of the peak in the 450\,$\micron$ dust continuum map of Phillips 
	et al. (2001). Finally, we list an upper limit for our X-ray non-detection of HH\,26-IR, the possible 
	driving source for the HH\,26 outflow (Schwartz et al. 1997, wherein the object is known as IRS 4). 
	The B1950 location for that source was taken from the discovery paper of Davis et al. (1997) and 
	precessed to an equinox of J2000.   	  

    \subsection{The Infrared Imaging}

	Two sets of {\it JHK} near-infrared images were obtained of the HH\,24--26 region on 18 November 
	2002, using the University of Hawaii QUIRC HgCdTe camera array (Hodapp et al. 1996). The instrument 
	was mounted at the f/10 Cassegrain focus of the 2.2-m telescope on Mauna Kea. The image scale was 
	0\farcs19 pixel$^{-1}$, the field of view was 3\farcm2. One set of images was centered on SSV\,63 
	and HH\,24, the other mid-way between SSV\,59 and 60 in the vicinity of HH\,25--26. At both 
	locations, a sequence of $3\times 10$ s exposures was obtained at each of 20 separate positions, 
	offset by 10\arcsec\ in right ascension and declination, to produce a heavily sampled dither-pattern.  
	The same cycle was repeated in each of the three filter bands.  We alternated the observations of 
	both HH fields with similar integrations on a pair of flanking fields 150\arcsec\ E and W, from 
	which the bright stars were subsequently removed in order to create suitable sky frames. 
	Flat-field frames were created by median-averaging observations of the 2.2-m dome under illumination 
	by incandescent lamps. In the reductions, the dithered HH frames were flat-fielded and sky-subtracted, 
	then aligned by cross-matching the positions of bright stars to form an enlarged median-averaged 
	mosaic. The cumulative exposure time in the center of the final co-added image in each filter is 
	$\sim500$ s. Taking into account the intermittent, non-photometric observing conditions, we estimate 
	brightness limits of 20.9 mags, 20.1 mags, and 18.5 mags in $J$, $H$, and $K$, respectively.  
	The typical seeing in the course of the observations was 0\farcs5 in $K$ (FWHM).  However, the PSF 
	of stellar images in the co-added data proved to be somewhat broader than this, and also slightly 
	distorted, as a result of a misalignment of the primary mirror of the 2.2-m telescope.

	Our images of HH\,24 in the three colors are shown in Fig. 4. The positions of the individual 
	components of SSV\,63, as listed in Table 2, are marked on the $J$-band image.  We have also 
	identified the expected locations of the two most prominent HH emission knots, 24A and 24B.  
	The position from Table 2 of the Class 0 object, HH\,24-MMS, is also noted. That source is 
	faintly visible in the $K$ image, the bandpass of which includes a number of H$_2$ lines, and is 
	also very weakly present in the $H$ image, whose bandpass includes [\ion{Fe}{2}] $1.64\,\micron$, 
	a low-excitation emission line that is observed in the jets of some HH objects (e.g., Reipurth 
	et al. 2000).  As noted earlier, the elongated jet  associated with HH\,24-MMS is very prominent 
	in the continuum-subtracted, narrow-band H$_2$ 
	images of Bontemps et al. (1996, their Fig. 1a) and Davis et al. (1997, their Fig. 2), and also 
	in the Fabry-Perot image of Davis et al. (2002, their Fig. 3). The N--S components of SSV\,63W are 
	cleanly separated in our $J$-band image, but are heavily exposed in the longer-wavelength images, 
	as is true of SSV\,63E.  An extended halo around the latter component, pointing westward, becomes 
	increasingly evident with increasing wavelength. Contrary to previous reports (Moneti \& Reipurth 
	1995; Terquem et al. 1999), we see no evidence to suggest this star is double or that the emission 
	peak in $H$ lies 1\arcsec\ west of the peak in $K$. In our images, there is no shift in the 
	position of the peak with color to a limit of 1 pixel or 0\farcs2. 

	The coordinates of the radio and X-ray source SSV\,63NE place it adjacent to two very red 
	patches of nebulosity known from earlier work (e.g., Zealey et al. 1992).  A false-color image 
	that combines the three separate panels of Fig. 4 exhibits a noticeable gradient in color from 
	W to E, possibly originating at SSV\,63E. The variation in color suggests that both patches of
	infrared nebulosity may be illuminated by an embedded source to the west, either the Class I source, 
	SSV\,63E, which Scarott et al. (1987) proposed as the illuminating source for the diffuse optical 
	nebulosity of HH\,24, or else a much more deeply embedded infrared object at the location of 63NE. 
	The radio and  X-ray coordinates of SSV\,63NE coincide to within 3\farcs3 with the position for the 
	predicted but unobserved ``Jet G Star'' of Jones et al. (1987), which those authors proposed to be 
	the driving source of the HH\,24G outflow (see also Mundt et al. 1991, in particular the highly detailed 
	[\ion{S}{2}] image in their Fig. 17). The absence in Fig. 4 of any obvious point source at the same 
	location to a limit of $K = 18.5$ mag implies that this star, presuming it does exist, lies exceedingly 
	deep within the molecular cloud. Given our detection of X-ray emission at the radio position, there 
	can be no doubt that SSV\,63NE, if it is a protostar of Class 0, has already evolved past the stage 
	of isothermal collapse.  Submillimeter maps of the region obtained with the SCUBA bolometer array 
	on the JCMT (Phillips et al. 2001; Mitchell et al. 2001) show extended emission at the position of
	SSV\,63 but are unable to resolve 63NE from the two other radio components, 63E and 63W. Consequently, 
	although it seems quite likely that SSV\,63NE is a bona fide YSO, its exact status is not yet entirely 
	settled.

    	Our infrared images of the southern region containing HH\,25 and 26 are presented in Fig. 5.                                         	Again, for orientation, the positions of the notable objects in the field of view are marked. The
	Class 0 protostar, HH\,25-MMS, is invisible in both the $J$ and $H$ images, although a hint of
	emission (presumably from ro-vibrational lines of H$_2$) can be seen at the radio position of this 
	object in the $K$-band image.  HH\,25A and SSV\,59 are both immersed in nebulosity. Nebulae of 
	similar appearance have been seen in high resolution images of other embedded protostars, e.g., 
	HL Tau (Close et al. 1997, especially their Figs. 2 and 4) and interpreted as the outline of bipolar 
	cavities. SSV\,59 has been classified as a Class I object. No star has been observed before at the 
	location of HH\,25A at optical wavelengths.  However, this HH knot is uncommonly bright in the $J$ 
	and $H$ frames in Fig. 5 to be merely the ``working surface'' of a high-velocity outflow from 
	HH\,25-MMS, located immediately to the south. Our underexposed 1.9--2.5\,$\micron$ infrared 
	spectrum of HH\,25A displays an array of very strong emission lines of H$_2$, which likely 
	arise from shock-heated gas, but no spectral signatures of an underlying star. 

	A very red, compact emission knot is observed at the location of HH\,25C, in approximate alignment 
	with HH\,25A, SSV\,60, and HH\,25D. As mentioned earlier, we believe SSV\,60 is a heavily reddened 
	background giant, not a member of L1630.  No other plausible candidate for the exciting source of 
	HH\,25 is revealed by our images.

    \subsection{The Infrared Spectroscopy}

 	Infrared spectra of SSV\,59, 60, 61, 63E, 63W (both the north and south components combined), 
	and HH\,26-IR were acquired with the SpeX cross-dispersed, medium-resolution spectrograph 
	(Rayner et al. 2003) at the 3.0-m NASA IRTF telescope on Mauna Kea on 3 January and 1--3 
	February 2003. SpeX covers the full wavelength range from 0.8 to 5.5\,$\micron$ in two grating 
	settings. We observed each source in the 1.9--4.2\,$\micron$ wavelength region, using the LXD
	setting; the two brightest objects, SSV\,60 and 61, were also observed in the 0.8--2.4\,$\micron$ 
	region with the SXD setting. Observing conditions were photometric and dry throughout, with t
	ypical seeing values near 0\farcs5 FWHM (measured in $K$-band). Slit widths of 0\farcs3 and 
	0\farcs5 at the short- and long-wavelength settings provided a nominal spectral resolving power, 
	$\lambda/\Delta\lambda$, of 2000 and 1500, respectively.  Our on-source integration times ranged 
	from 300 to 1200\,s and resulted in signal-to-noise ratios of 20--50 in both the $K$ and $L$ bands.  
	
	Each observing sequence began with a nearby (less than 0.05 difference in airmass) A0\,V 
	standard star, moved to the target object, and finished with a set of internal calibration 
	spectra. To cancel the sky background, we observed in the conventional way by nodding 
	along the 15\arcsec-long  slit. The data were reduced with the  
	{\sc spextool}\footnote[5]{http://irtfweb.ifa.hawaii.edu/Facility/spex/} software package (see 
	Cushing et al. 2003 for details). Spectra were extracted using a 1\farcs2 aperture along the 
	slit, corrected for telluric features, and calibrated in flux by means of the standard star spectra 
	according to the techniques that have been described by Vacca, Cushing, \& Rayner (2003). 
	The processed spectra are shown in Fig. 6, where they are plotted on a logarithmic flux scale.

	The infrared continua of SSV\,60 and 61 come to a peak blueward of the $K$-band, near 1.9 
	and 1.5\,$\micron$, respectively. Absorption lines from a variety of neutral and ionized metals, 
	including \ion{Ca}{1}, \ion{Fe}{1}, \ion{Mg}{2}, \ion{Al}{1}, \ion{Ti}{1}, and \ion{Na}{1}, can 
	be seen in the spectra of both stars. No emission lines are observed in the spectrum of SSV\,60. 
	The Br\,$\alpha$ line of \ion{H}{1} at 4.05\,$\micron$ appears in emission in the spectrum of 
	SSV\,61; its equivalent width is entirely consistent with the observed strength of H$\alpha$, 
	Case B recombination, and very little foreground reddening. Both spectra exhibit deep absorption 
	bands from the first overtone ro-vibrational transitions of CO near 2.3--2.5\,$\micron$. Assuming 
	the observed 
	atomic and molecular features are photospheric in origin, we have used their relative strengths 
	to estimate the spectral type and luminosity class of the underlying stars in the manner described 
	by Kleinmann \& Hall (1986), Hodapp \& Deane (1993), and Greene \& Meyer (1995). Here we have 
	adopted  Aspin's (2003) formulation, in which an atomic index formed by the sum of the equivalent 
	widths of  the \ion{Na}{1} doublet at 2.21\,$\micron$ and the \ion{Ca}{1} triplet at 2.26\,$\micron$
	is compared to a molecular index represented by the equivalent width of the $v$=2--0 CO bandhead 
	at  2.29\,$\micron$ (see Fig. 7). According to their atomic and molecular index values,  SSV\,60 
	(4.2\,\AA, 10.0\,\AA) is definitely a giant, whereas SSV\,61 (5.9\,\AA, 9.4\,\AA) is intermediate 
	to a dwarf and giant (a not uncommon finding for a TTS). We have combined the spectral indices, the 
	strengths of the various photospheric absorption lines of the metals, the deep CO bandheads, and 
	the general continuum shapes to estimate spectral classifications of K5--M0\,III for SSV\,60 and 
	M0--M2\,III/V for SSV\,61. Those spectral types, resting solely on the infrared spectra, are consistent 
	with the locations of both stars in a ($J-H$, $H-K$) color-color plot based on their 2MASS photometry, 
	and also with the effective temperatures we have derived following the procedure of Greene \& 
	Meyer (1995), which makes use not only of the absorption line strengths but the broadband 
	colors as well. The latter technique provides, among other things, an estimate of the fraction of  
	excess light from veiling continuum emission, which we find to be constrained to low values for both 
	stars ($r_H$, $r_K < 0.15$ at $H$ and $K$).  
	 
	The association of SSV\,61 with the L1630 molecular cloud is hardly in question, but the same 
	cannot be said for SSV\,60. With its bright, attendant nebula, SSV\,61 would be a rare star indeed 
	to be in the far foreground of this region, while the absence of any ice absorption at 3.08\,$\micron$ 
	in its infrared spectrum (Fig. 6) argues against the likelihood it lies far  behind the molecular 
	cloud at a very considerable distance from Earth. For an early M giant, SSV\,60 is clearly much 
	more heavily absorbed than SSV\,61, as is apparent from its  spectral energy distribution shortward  
	of 2\,$\micron$ and its strong ice absorption band. The broad ice feature is common to the spectra 
	of other sources which are either background to or deeply embedded within molecular clouds, where 
	the ice has condensed onto dust grains but not undergone significant processing at high temperature  
	(Smith, Sellgren, \& Tokunaga 1989). For SSV\,60, the optical depth at band center, $\tau_{ice}$ 
	=\,0.96, leads (according to the results of Whittet et al. 1988) to an optical extinction of roughly 
	$A_v \approx 17$ mag. Such heavy extinction is not to be found in the foreground to L1630.  On the 
	other hand, SSV\,60 shows none of the signposts of youth or activity that are expected for an embedded 
	Class I protostar---no X-ray emission, no high velocity outflow or jet in CO or H$_2$, no far infrared 
	or sub-mm emission. We conclude, therefore, that SSV\,60 is a background giant star that is unconnected 
	either with L1630 or with HH\,25, on which it is apparently coincidentally projected. 

	The strong red continua and deep ice bands of HH\,26-IR, SSV\,59, 63E, and 63W, on the other 
	hand, suggest that each of those objects is deeply embedded within the L1630 cloud. Their observed 
	ice-band optical depths range from $\tau_{ice} = 0.8$ (for HH\,26-IR) to $\tau_{ice} = 1.3$ (for 
	SSV\,63E and W), corresponding to approximate $A_v$ values of 10 to 20 mags.  At close inspection, 
	the ice band spectra of SSV\,59 and 63E also show a broad, shallow depression near 3.45\,$\micron$. 
	That feature, according to Smith et al. (1989), can be identified with the C--H stretching modes of 
	hydrocarbons  (in the CH$_2$ and CH$_3$ subgroups). Emission lines of \ion{H}{1} are visible in the 
	spectra of  all four objects, including Br\,$\alpha$ 4.05\,$\micron$, Br\,$\gamma$ 2.17\,$\micron$, 
	and (in the case of SSV\,63E) $n$=8--5 Pf\,$\gamma$ 3.74\,$\micron$.  The emission is a clear 
	indication of ionized gas in the close vicinity of each star, but whether the warm gas is inflowing
	or outflowing is a matter of continuing debate (e.g., Najita, Carr, \& Tokunaga 1996; Folha \& 
	Emerson 2001).  There are no signs in our spectra of any photospheric absorption lines, which we 
	suspect may be hidden beneath a large continuum veiling flux. Emission lines from ro-vibrational 
	transitions of H$_2$ are present throughout the $K$-band spectra of  SSV\,63W and HH\,26-IR, the 
	2.12\,$\micron$ $1-0\,S$(1) line being the strongest (Fig. 8). The same lines are often observed 
	in the spectra of Class I objects (Greene \& Lada 1996). They are also commonly observed in the 
	spectra of high velocity outflows (e.g., Gredel 1994, 1996) and have been attributed to excited 
	H$_2$ molecules that form behind shocks which arise as jet material collides with the ambient gas 
	in a cloud (Garden, Russell, \& Burton 1990). The H$_2$ emission in our spectral images does not 
	appear to be spatially extended in a N--S direction along the slit. However, it has already been 
	shown by Davis et al. (2002) that the H$_2$ emission of HH\,26-IR has a modest extension along 
	its jet in the NE--SW direction. 

	The $K$-band spectra of SSV\,59, 63E, and HH\,26-IR (Fig. 8) display the first overtone ro-vibrational
	CO bands in emission. Various explanations have been offered for the CO lines: they may be formed 
	in the innermost portion of a Keplerian disk or in the neutral component of a stellar wind or outflow 
	(Carr 1989; Chandler, Carlstrom, \& Scoville 1995a; Najita et al. 1996).  Alternatively, the emission 
	may be produced by accretion of disk material onto the central star via a magnetic ``funnel flow'' 
	(Martin 1997; see, however, Najita et al. 2003).  In any case, Biscaya et al. (1997) have
	demonstrated that in some cases the CO bands can be highly variable, perhaps even exhibiting periodic 	behavior. Carr (1989), in fact, observed the CO bands of SSV\,59 in absorption, whereas they are 
	definitely in emission in our SpeX spectra, as can be readily seen from Fig. 8.

    \section{DISCUSSION}
    
    	\subsection{The Class 0 Sources}

	We detected X rays from none of the three Class 0 YSOs within the {\it Chandra} field of view.  
	Two of those embedded objects, HH\,24-MMS and HH\,25-MMS, are continuum radio sources 
	and therefore must have a component of ionized gas; the third object, LBS\,17-H, has not been 
	detected at radio wavelengths and hence may be in an earlier stage of evolution than the others, 
	or else its radio emission is much more optically thick and therefore below current detection
	levels (Gibb 1999). The radio spectral indices of 24-MMS and 25-MMS are slightly positive (Anglada 
	et al.  1998).  The indices are compatible with thermal free-free radiation (Reynolds 1986), but 
	do not  definitively rule out a possible non-thermal gyrosynchrotron origin. We note that the 
	spectral index for gyrosynchrotron emission is most often strongly negative but it can also be 
	weakly positive, as for a thermal source (e.g., G\"{u}del 2002). In the case of HH\,24-MMS, the 
	radio source is spatially resolved (Bontemps et al. 1996; Reipurth et al. 2002).  Therefore, it 
	is almost certainly free-free emission from a warm ionized jet, although some portion of it may 
	be thermal emission from dust  (Ward-Thompson et al. 1995).    
	
	We can suggest several plausible explanations to account for the dearth of X rays from the Class~0 
	objects in L1630: (a) assuming the three YSOs are in the main phase of protostellar accretion, 
	they may not yet have developed accretion shocks that are strong enough to heat infalling gas to 
	the million degree temperatures that are needed to generate X rays, or else their magnetic fields 
	may not be strong enough or sufficiently well organized on large scales to produce intense 
	reconnection heating of the infalling gas; (b) alternatively, if hot gas is now present around the 
	star, the temperature of the gas may be sufficiently low (below 1--2 MK) that it emits only a very 
	soft spectrum of X-ray photons, which is unable to penetrate the dense envelope that surrounds the 
	growing protostar; or (c) if there is X-ray emitting gas now present at temperatures that are high 
	enough to form hard X~rays with energies of 2--10 keV, the extinction in the surrounding envelope 
	and cloud may be sufficiently large to absorb most (or all) of the X-ray photons and to depress 
	the emergent radiation below levels that {\it Chandra} can detect.  

	Here we address the last possibility. Since, by definition, Class 0 protostars are especially 
	faint at near-infrared wavelengths, we do not have $JHK$ colors or IRTF spectra for the Class~0 
	YSOs in L1630 from which we might estimate the foreground extinction to those objects. Instead, 
	we have estimated the line-of-sight extinction to each source by making use of the relationship 
	between visible extinction and 850\,$\micron$ brightness given by Mitchell et al. (2001), $A_v = 
	77 S_{\nu}$. Applying that expression in turn to the submillimeter data for HH\,24-MMS, HH\,25-MMS, 
	and LBS\,17-H, we find $A_v$ estimates  of 200, 100, and 86 mags, respectively. In each case, the 
	derived extinction is larger than the value of 48 mags we previously adopted from our {\sc xspec} 
	model of the {\it Chandra} spectrum of SSV\,63W.  

	Given the foregoing extinction estimates, we have used the HEASARC WebPIMMS tool to predict a 
	{\it Chandra} detection limit for each Class 0 YSO, computing for each one the intrinsic X-ray 
	luminosity that yields the observed $2\sigma$ count rate limit listed in Table 2. In our 
	simulations, we have adopted a single-temperature thermal model for the X-ray emission and assumed 
	a normal value of the gas-to-dust ratio in order to convert the submillimeter-based $A_v$ into a 
	column density of N$_H$, following Bohlin et al. (1978). Guided by our {\sc xspec} calculations, 
	we have considered two different source temperatures which should bracket an adequate range of 
	spectral hardness, $kT = 5$ keV and $kT = 2$ keV.  In the high-temperature case, the tabulated count 
	rate limits for HH\,24-MMS, HH\,25-MMS, and LBS\,17-H correspond to $2\sigma$ detection limits in 
	$\log L_x$ of 30.30, 30.00, and 29.75, respectively. The limits in $\log L_x$ for the low-temperature 
	models are 0.3 dex higher. For both temperatures, the predicted luminosity limits are far less than 
	the intrinsic $L_x$ we derived from {\sc xspec} models of the Class~I object SSV\,63W, which we 
	detected at $\log L_x = 31.16$. To have escaped detection at that $L_x$, i.e., to have resulted in
	fewer than $\sim25$ net X-ray counts in our exposure time of 63 ks, the Class 0 objects would each 
	have to suffer more than 400 mags of extinction (or else be intrinsically very much softer X-ray 
	sources than we have assumed). Therefore, to a very high significance level, it appears all three 
	Class~0 objects are intrinsically much fainter in X rays than 63W, even when the latter source is 
	quiescent rather than flaring. The X-ray luminosity of the eastern component of SSV\,63, the Class~I 
	YSO 63E, which we also detected with {\it Chandra}, is comparable in strength to the luminosity upper 
	limits for the Class~0 objects.  Thus, we can further assert that the Class~0 sources are fainter than 
	63E at the $2\sigma$ confidence level. On the other hand, because the two other established Class~I 
	objects in the same observation, SSV\,59 and HH\,26-IR, were not detected by {\it Chandra}, the X-ray 
	brightness of individual Class~I sources must have a wide range of values, even within the same star 
	forming region. {\it Chandra} observations of Class I YSOs elsewhere in Orion (Tsujimoto et al. 2002) 
	and in Ophiuchus (Imanishi et al. 2001) bear that point out.  Thus, although the Class~0 objects 
	in L1630 are clearly fainter than {\it some} of the Class~I objects in the same cloud, they may not 
	be fainter than {\it all} of them.  In terms of an evolutionary trend, until more is known about 
	the lower range of $L_x$ values for Class~I objects, and the extent to which such objects are 
	variable, it is premature to draw any firm conclusions about whether the Class~0 sources as a group 
	are systematically less luminous in X rays than are the more advanced Class~I sources which are their 
	descendants.

	\subsection{The Class I Sources}

	The submillimeter map of Mitchell et al. (2001) leads to an $A_v$ estimate (47 mags) for the 
	combined (spatially unresolved) SSV\,63EW that is entirely in agreement with the extinction we 
	have derived from our {\sc xspec} modelling of the {\it Chandra} spectrum of 63W.  Both the E 
	and W infrared components of SSV\,63 were detected in X rays; they are also both thermal radio 
	emitters (Anglada et al. 1998). HH 26-IR was not detected in X rays, but may be a weak radio source 
	according to Gibb (1999). SSV\,59 is neither an X-ray source nor a radio source.  However, the 
	near-infrared spectrum of SSV\,59, as well as that of HH\,26-IR (Fig. 6), exhibits emission in the 
	Brackett lines of hydrogen, implying that both objects are associated with ionized gas. The presence 
	of a resolved H$_2$ jet appended to HH 26-IR (Davis et al. 2002) suggests that its radio emission 
	may originate from the jet and be thermal in nature; the radio spectrum of SSV\,59 may be optically 
	thick and thus undetectable. It is perhaps significant that the HH filaments and knots identified 
	with the northern Class~I YSOs (the various alphabetical components of HH\,24) are much more highly 
	developed than the ones in the south (HH\,25--25).  This is perhaps an indication that the southern 
	region is relatively younger and has not yet reached the age where its embedded YSOs have formed 
	strong radio sources or possess the million degree plasma required to produce X rays.

	\subsection{SSV\,60 and SSV\,61}

	The weak-line T Tauri star SSV\,61 is by far the brightest source in our {\it Chandra} image.  
	It is also a highly variable radio source, which suggests a non-thermal origin for its radio flux 
	(Anglada et al. 1998; Gibb 1999). As we noted in \S{2.3}, there is very little chance of our
	having greatly overestimated the distance and consequently the $L_x$ of this star. For an assumed 
	distance of $d = 450$ pc, the radio and X-ray luminosities of SSV\,61 closely fit the $L_R-L_x$ 
	relation for normal late-type giants, lying among the tidally locked binaries at the top end of 
	the activity range (Drake, Simon, \& Linsky 1989; Dempsey et al. 1993). This may be an indication 
	that SSV\,61 is itself a close binary. In any event, it is safe to say that the emissions of this 
	star at radio as well as X-ray wavelengths are coronal in origin.  
	
	The near-infrared spectrum and 2MASS colors of SSV\,61 in a  ($J-H$, $H-K$) two-color plot are 
	consistent with those of a lightly reddened and lightly veiled early M giant star, for which 
	$A_v < 2$ mags.  However, the observed infrared magnitudes, combined with the bolometric correction 
	for an early M giant ($K = 8.36$, B.C.$_K$ = 2.60 mags), imply a very high luminosity, $L_{\rm bol}
	\sim 7.5 L_{\odot}$, which places the star well above the stellar birthline at a logarithmic effective 
	temperature of $\log T_{\rm eff}$ = 3.5--3.6 (Stahler 1983; Mercer-Smith, Cameron, \& Epstein 1984).
	At that $L_{\rm bol}$, the ratio of the X-ray to the bolometric luminosity of SSV\,61 is at the 
	saturation limit for the coronal X-ray emission of late-type stars, $L_x/L_{\rm bol} \lesssim 10^{-3}$. 
	That maximum X-ray brightness limit is well established from observations of many different kinds of 
	late-type stars, among them wTTS like SSV\,61, and so any substantial reduction in our $L_{\rm bol}$ 
	estimate for SSV\,61 would make it all the more difficult to understand the origin of its intense 
	X-ray brightness. On the other hand, as one might surmise from the optical polarimetry of Scarrott 
	et al. (1987), the 2MASS photometry of SSV\,61 may be heavily contaminated by scattered light from 
	the nebulosity that is present on all sides of the star. A large proportion of the near-infrared 
	light of the more deeply embedded Class~I objects is thought to be contributed by scattering in 
	the protostellar envelope (Kenyon et al. 1993).  In that instance the scattering adds no continuum 
	veiling to the spectrum of the central star itself (Calvet et al. 1997). The same scattering 
	mechanism, applied to SSV\,61, would therefore be consistent with the low veiling ratio we have 
	determined from our SpeX observation of SSV\,61.  
	
	In the submillimeter map of Mitchell et al. (2001), there is an 850$\micron$ peak that coincides in
	position with SSV\,61.  The value of $A_v$ implied by the submillimeter flux is 40 mags or more, 
	which is inconsistent with the near absence of reddening established by the infrared colors and 
	ice-band spectrum of SSV\,61.  This suggests that the cold dust producing the 850$\micron$ peak is 
	most likely located behind SSV\,61 and that the star sits in front of the cloud, possibly 
	illuminating its front face.                                                                                                                                                             

	In the infrared color-color plot, the 2MASS colors of SSV\,60 place this star on the reddening 
	line for late-type giants, consistent with the spectral classification we obtained from our IRTF 
	spectrum. However, the bright near-infrared magnitudes of this star ($K = 8.44$), coupled with 
	the large extinction implied by the deep 3$\micron$ ice band we have observed ($A_v \approx 17$), 
	cannot be reconciled with membership in L1630.  The best match we can find for the spectral type, 
	the ($J -H$, $H -K$) colors, the large amount of reddening, and the near-infrared magnitudes are 
	for an early M background giant, which is located beyond the L1630 cloud at a distance of $d = 2.0$ 
	kpc and with a luminosity of $L_{\rm bol} = 600 L_{\odot}$.

    \section{CONCLUSIONS}

	We have acquired a deep X-ray image of the HH\,24--26 region in the L1630 cloud with the ACIS-S 
	camera on {\it Chandra}, which has resulted in the detection of a number of anonymous 2MASS infrared 
	sources and optically visible H$\alpha$ emission-line stars, including the X-ray luminous coronal 
	source, the wTTS SSV\,61. Also detected at what appear to be their quiescent levels of activity are 
	two deeply embedded Class~I protostars (SSV\,63E and W), which have been identified in earlier 
	work as the source of energetic molecular and ionized flows in the region. The two were blended
	together in an earlier detection by ASCA, but are well separated at the much higher spatial 
	resolution of {\it Chandra}. Two other HH-associated Class~I objects (SSV\,59 and HH\,26-IR) as 
	well as three Class~0 YSOs in the same cloud (HH\,24-MMS, HH\,25-MMS, and LBS 17-H) were not detected.  
	Although X-ray observations of several different star forming regions thus far reveal a striking lack 
	of Class~0 detections, no firm conclusions can be drawn from our small sample of YSOs regarding a 
	possible evolutionary trend in the intrinsic X-ray luminosities of the Class~0 and Class~I protostars. 
	X-ray emission was detected from a deeply embedded continuum radio source (HH\,24NE), whose location 
	within HH\,24 suggests that it may be the highly obscured driving source for one of the long HH filaments 
	in this area. The object has no visible or near-infrared counterpart, and therefore is potentially 
	a Class~0 or extreme Class~I object.\\

	This research has made use of the {\sc simbad} database, operated at {\it CDS},
	Strasbourg, France, and is based in part on observations made with the \chandra.  
	Support for this work was provided by \nas\ through Chandra Award Number GO2-3007X   
	issued by the \chandra\ Center, which is operated by the Smithsonian Astrophysical 
	Observatory for and on behalf of \nas\ under contract NAS8-39073. The Digitized Sky 
	Surveys (DSS) were produced at the Space Telescope Science Institute under NASA Grant 
	NAG W-2166 and are based on photographic data obtained using the Oschin Schmidt 
	Telescope on Palomar Mountain and the UK Schmidt Telescope with the permission of those 
	institutions. The Guide Star Catalogue-II is a joint project of the Space Telescope 
	Science Institute and the Osservatorio Astronomico di Torino. Space Telescope Science 
	Institute is operated by the Association of Universities for Research in Astronomy, 
	for the National Aeronautics and Space Administration under contract NAS5-26555. The 
	participation of the Osservatorio Astronomico di Torino is supported by the Italian 
	Council for Research in Astronomy. Additional support is provided by  European Southern 
	Observatory, Space Telescope European Coordinating Facility, the International GEMINI 
	project and the European Space Agency Astrophysics Division.  The Two Micron All Sky
	Survey (2MASS) is a joint project of the University of Massachusetts and the Infrared
	Processing and Analysis Center (IPAC)/California Institute of Technology, funded by
	the National Aeronautics and Space Administration and the National Science Foundation. 



    \newpage

	\figcaption{The field of view observed in X rays with {\it Chandra}. Shown here is the
	red image from the Digitized Sky Survey of the STScI.  The orientation and sky coverage
	of the ACIS-S detector chips S0--S4 are shown by the overlay.  The back-illuminated
	chips are S1 and S3. The boresight of the observation is near HH\,24 on chip S3.
	Other HH objects in the field are denoted by the diamond symbols. The locations of
	the X-ray sources are marked by circles, whose diameters are greatly exaggerated 
	here for visibility.     }  

	\figcaption{Image in the hard X-ray band (2--8 keV) of the area around SSV\,63. The 
	locations of the three VLA radio components of Reipurth et al. (2002) are indicated 
	by crosses. }

	\figcaption{X-ray spectrum of SSV\,61 observed with the ACIS-S3 detector on {\it Chandra}
	is shown in the top panel. A two-temperature thermal model computed with the APEC plasma
	emission code and folded through the up-dated ACIS detector response is plotted as a histogram
	curve. Channels in the observed spectrum have been merged to ensure a minimum of 20 counts 
	per bin. The lower panel shows the contributions of the residuals in each bin to the value 
	of $\chi^2$.}  

	\figcaption{QUIRC near-infrared images of HH\,24. ({\it left}) $J$-band, ({\it middle})
	$H$-band, and ({\it right}) $K$-band.  The angular scale, centering, and orientation 
	are identical for all three panels. Following the usual convention, north is up, east 
	is toward the left.  Our estimated $5\sigma$ photometric limits are 20.9 mag in $J$, 
	20.1 mag in $H$, and 18.5 mag in $K$.}
  
	\figcaption{QUIRC near-infrared images of HH\,25--26. ({\it left}) $J$-band, ({\it middle})
	$H$-band, and ({\it right}) $K$-band. North is up, east is to the left. Photometric limits
	are as in the previous figure.}
  
	\figcaption{The near-infrared (0.8--4.1\,$\micron$) spectra of six stars in L1630, observed 
	with the SpeX facility instrument on the IRTF.  }

	\figcaption{SpeX $K$-band spectra of SSV\,60 (top curve) and SSV\,61 (bottom curve), 
	identifying the alkali lines and the 2--0 CO bandhead used for spectral classification. 
	Both spectra have been normalized to the continuum and are offset for clarity.}

	\figcaption{SpeX $K$-band spectra of four embedded stars in L1630.  They are, in order from 
	top to bottom: SSV\,63W (both components combined), SSV\,59, SSV\,63E, and HH\,26-IR.  The 
	observed fluxes have been normalized to the continuum and the spectra offset for clarity.}

\newpage
\plotone{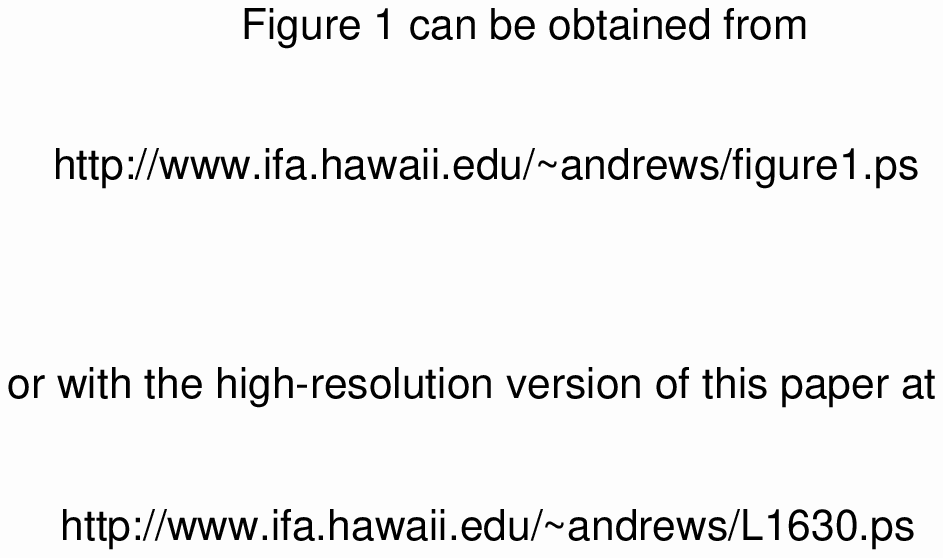}
\newpage
\plotone{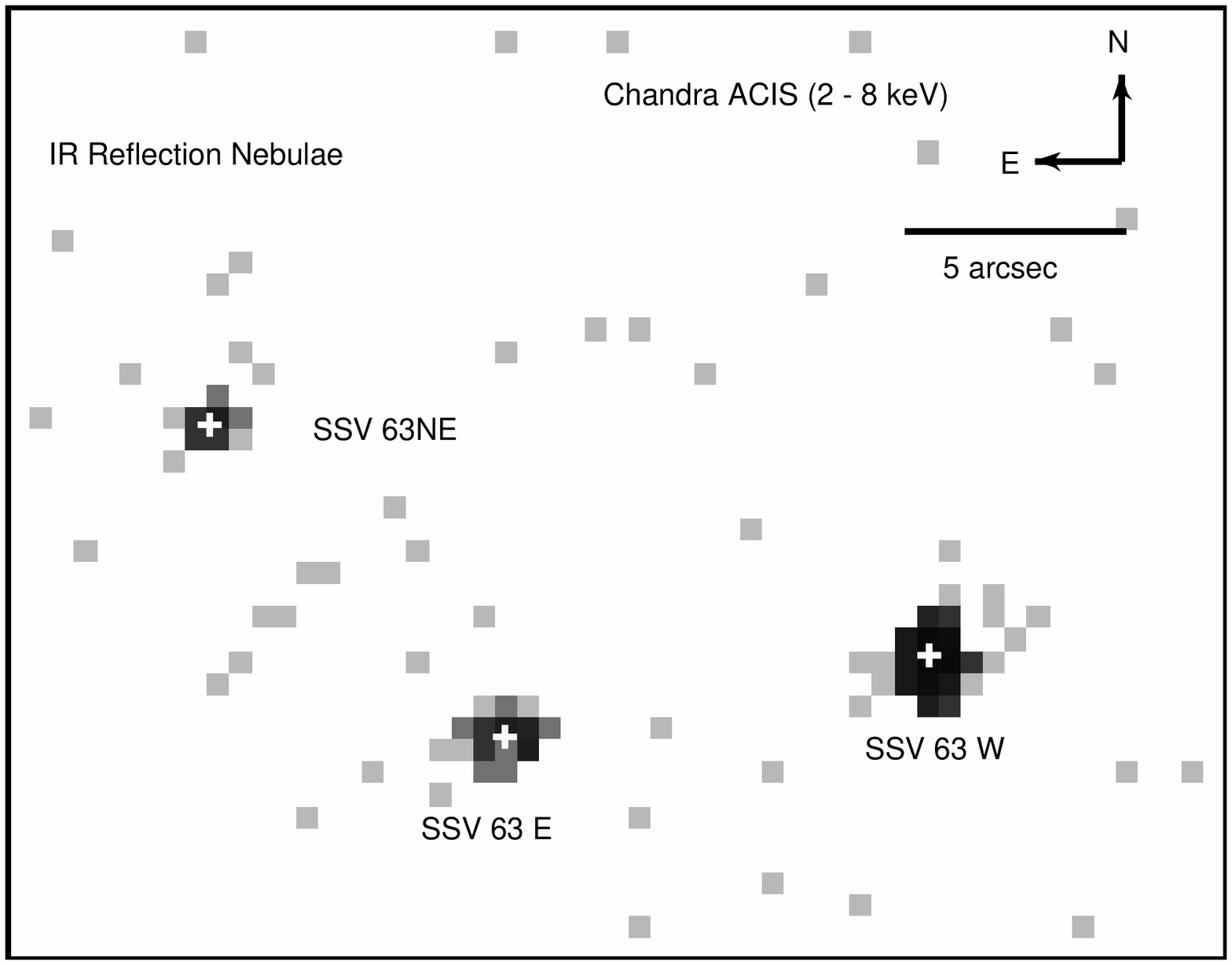}
\newpage
\epsscale{0.7}
\plotone{figure3.eps}
\newpage
\epsscale{1}
\plotone{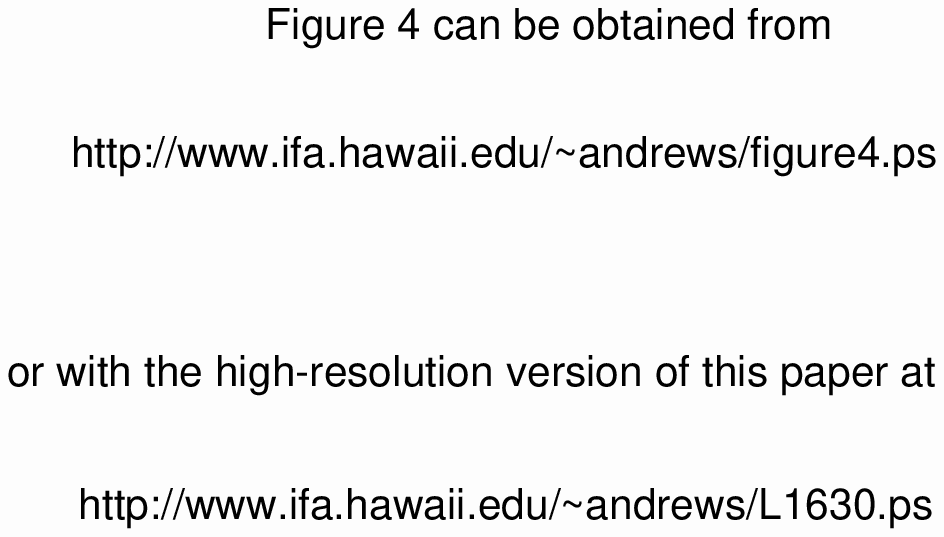}
\newpage
\epsscale{1}
\plotone{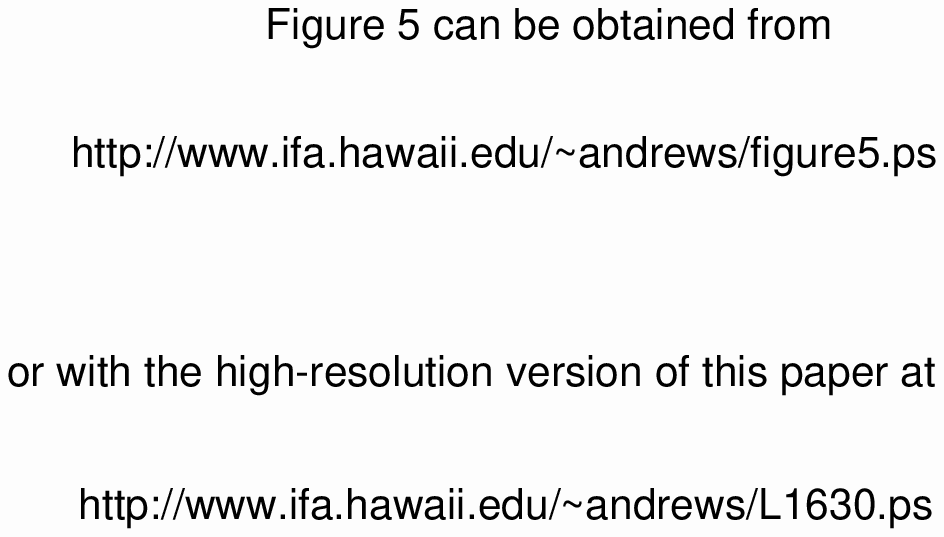}
\newpage
\epsscale{0.8}
\plotone{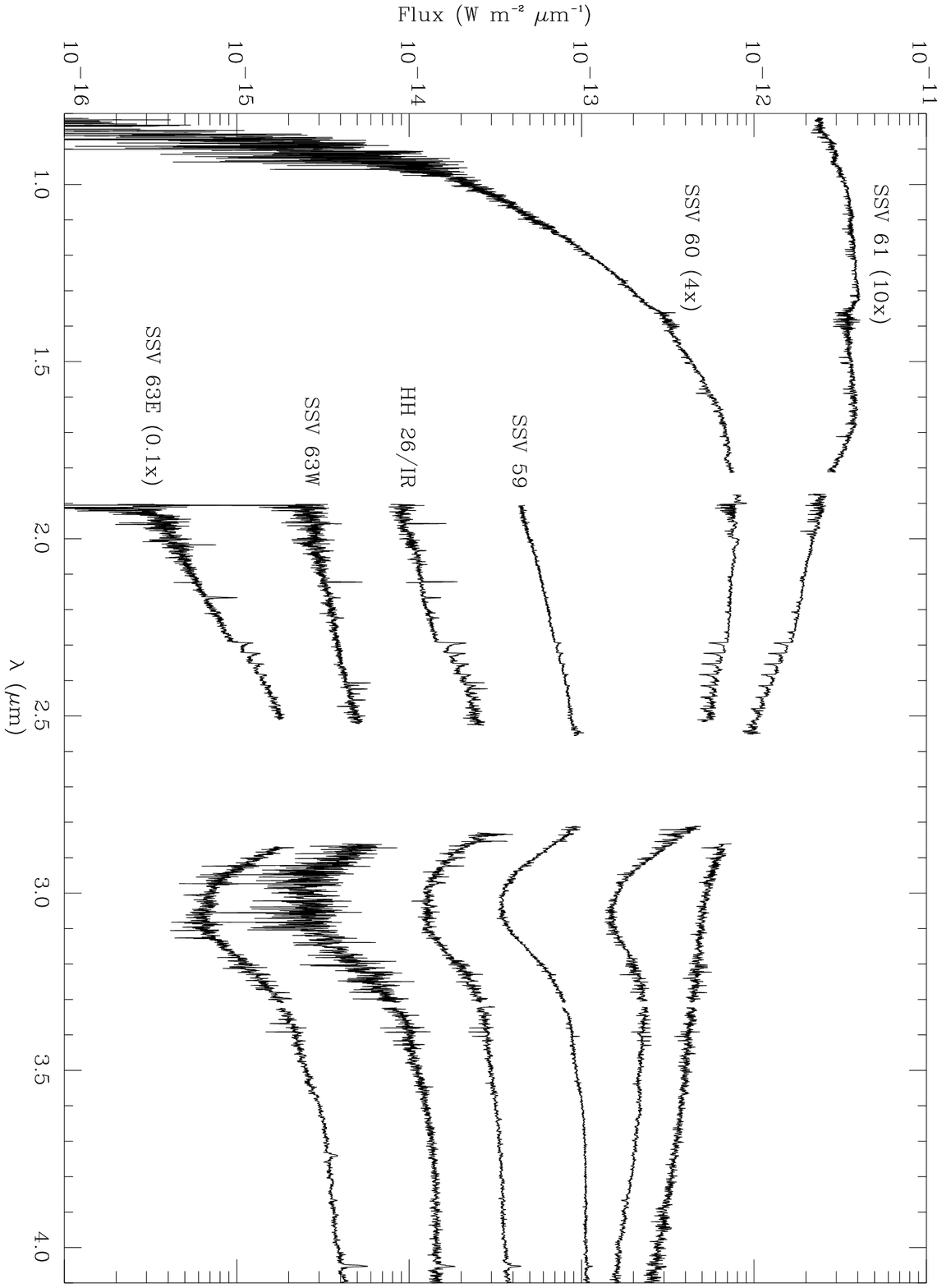}
\newpage
\epsscale{0.8}
\plotone{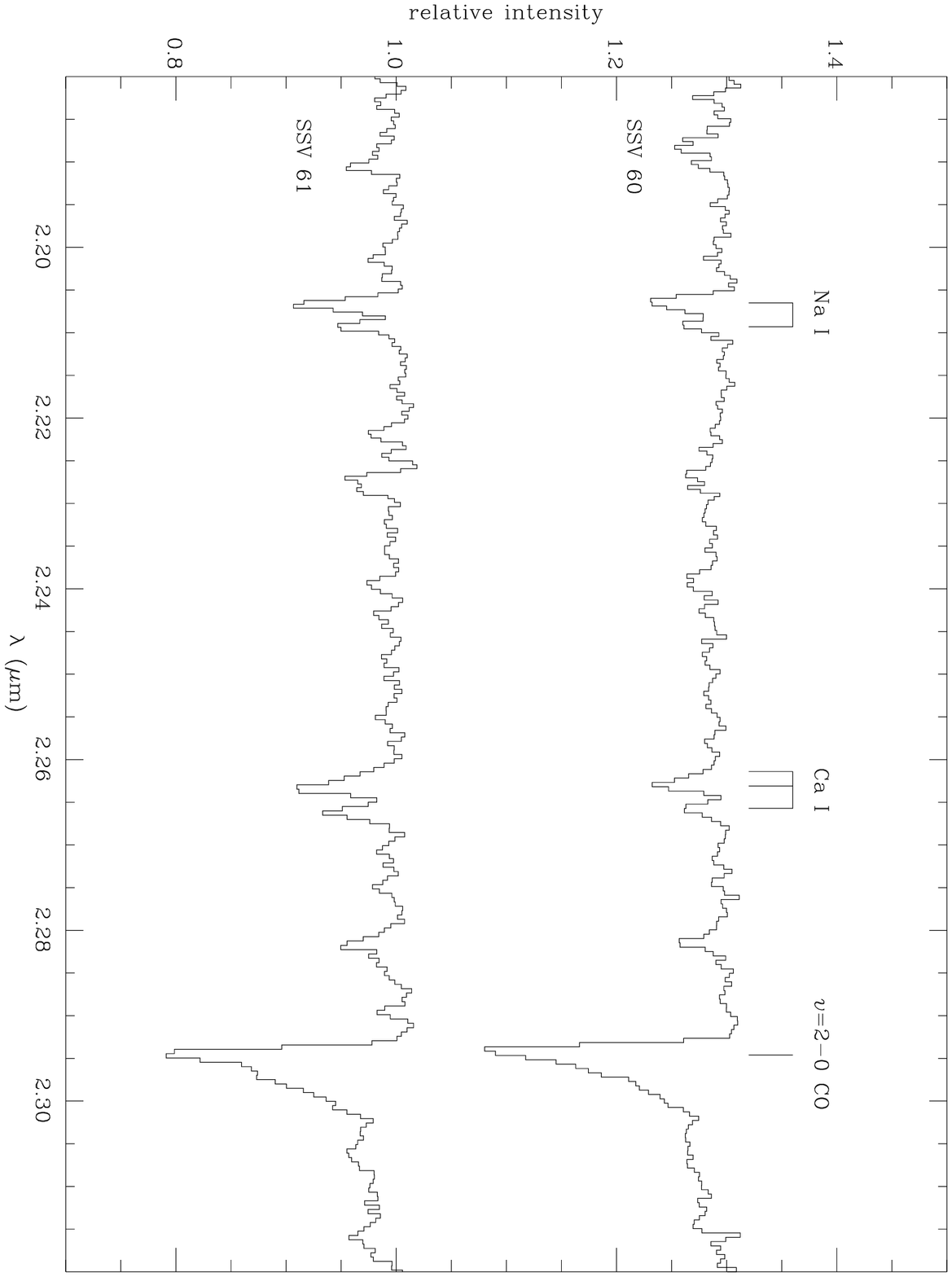}
\newpage
\epsscale{0.8}
\plotone{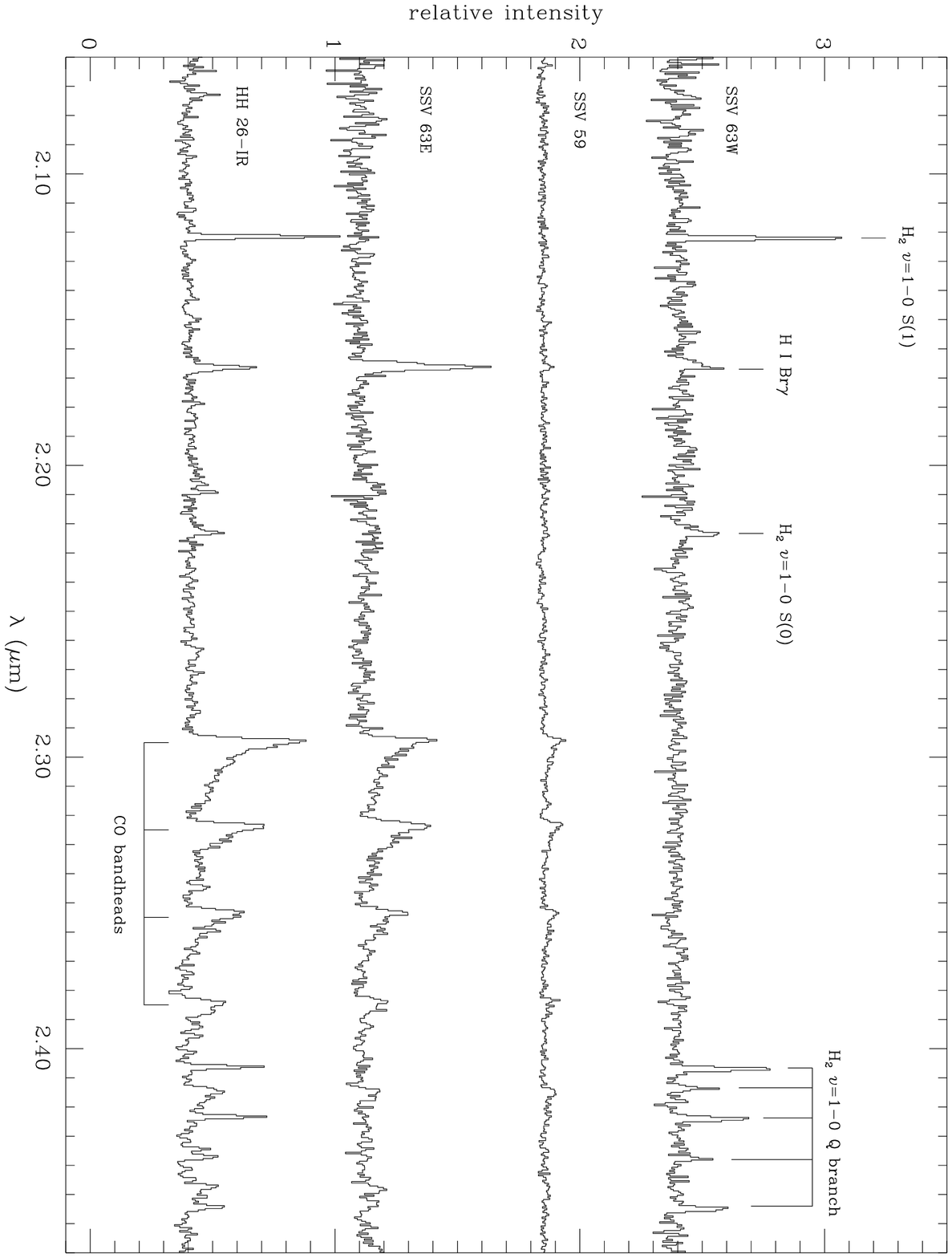}


\clearpage

\begin{deluxetable}{lccrlrr}
\tablenum{1}
\tablewidth{0pc}
\tablecolumns{7}
\tabletypesize{\small}
\tablecaption{X$-$Ray Detections in L1630}
\tablehead{ 
\colhead{}                        & \colhead{R.A.\tablenotemark{a}}              &
\colhead{Decl.\tablenotemark{a}}  & \colhead{$\theta$\tablenotemark{b}}          &
\colhead{}                        & \colhead{}                                   &
\colhead{Count Rate\tablenotemark{d}}		    
\\
\colhead{Source}                  & \colhead{(J2000)}                 &
\colhead{(J2000)}                 & \colhead{(\arcmin)}               &
\colhead{Source Identification}   & \colhead{Counts\tablenotemark{c}} & 
\colhead{(counts ks$^{-1}$)}
}
\startdata
1\,...... &  05 45 46.3  &  $-00$ 15 15  &  8.5  &  \nodata  &  $143\pm15$\phn  &  $2.81\pm0.30$  \\
2\,...... &  05 45 51.7  &  $-00$ 16 33  &  8.5  &  \nodata  &  $136\pm15$\phn  &  $2.66\pm0.30$  \\
3\,...... &  05 45 53.1  &  $-00$ 13 26  &  6.1  &  GSC2 S02000123390 (=KH$\alpha$\,174) &  $92\pm10$\phn  &  $1.55\pm0.17$ \\
4\,...... &  05 45 57.8  &  $-00$ 09 29  &  3.8  &  \nodata\tablenotemark{e}  &  $ 94\pm10$\phn  &  $1.54\pm0.16$  \\
5\,...... &  05 46 00.3  &  $-00$ 08 26  &  3.5  &  SSV\,65   &  $474\pm23$\phn  &  $8.68\pm0.42$  \\
6\,...... &  05 46 04.2  &  $-00$ 05 59  &  4.5  &  2MASS J05460427-0005591  &  $43\pm7$\phn\phn  &  $0.78\pm0.13$  \\
7\,...... &  05 46 07.8  &  $-00$ 10 01  &  1.3  &  SSV\,63W  &  $167\pm15$\phn  &  $2.65\pm0.24$  \\
8\,...... &  05 46 07.8  &  $-00$ 11 57  &  2.4  &  SSV\,61   &  $12960\pm115$  &  $251.92\pm2.23$  \\
9\,...... &  05 46 08.1  &  $-00$ 05 32  &  4.6  &  \nodata  &  $20\pm6$\phn\phn   &  $0.36\pm0.10$  \\
10\,....  &  05 46 08.5  &  $-00$ 10 03  &  1.2  &  SSV\,63E  &  $38\pm7$\phn\phn  &  $0.62\pm0.11$  \\
11\,....  &  05 46 08.9  &  $-00$ 09 56  &  1.0  &  SSV\,63NE  &  $24\pm6$\phn\phn  &  $0.39\pm0.10$  \\
12\,....  &  05 46 09.5  &  $-00$ 03 31  &  6.5  &  2MASS J05460960-0003312  &  $58\pm8$\phn\phn  &  $1.10\pm0.15$  \\
13\,....  &  05 46 09.8  &  $-00$ 05 59  &  4.0  &  2MASS J05460984-0005591  &  $29\pm5$\phn\phn  &  $0.52\pm0.10$  \\
14\,....  &  05 46 11.6  &  $-00$ 02 20  &  7.6  &  GSC2 S02000125079  &  $498\pm23$\phn  &  $9.69\pm0.44$  \\
15\,....  &  05 46 11.6  &  $-00$ 06 28  &  3.5  &  2MASS J05461162-0006279  &  $96\pm11$\phn  &  $1.70\pm0.19$  \\
16\,....  &  05 46 12.3  &  $-00$ 08 08  &  1.8  &  2MASS J05461226-0008078  &  $57\pm8$\phn\phn  &  $0.94\pm0.13$  \\
17\,....  &  05 46 13.0  &  $-00$ 18 27  &  8.5  &  \nodata  &  $277\pm18$\tablenotemark{f} \phn  &  $6.76\pm0.44$  \\
18\,....  &  05 46 16.3  &  $-00$ 01 19  &  8.7  &  \nodata      &  $68\pm9$\phn\phn  &  $1.34\pm0.18$  \\
19\,....  &  05 46 17.7  &  $-00$ 16 12  &  6.4  &  \nodata  &  $67\pm10$\tablenotemark{f} \phn   &  $1.19\pm0.18$  \\
20\,....  &  05 46 18.9  &  $-00$ 05 38  &  4.6  &  GSC2 S020001258   &  $1820\pm43$\phn  &  $32.81\pm0.78$  \\
21\,....  &  05 46 19.5  &  $-00$ 05 20  &  4.9  &  SSV\,64 (=LkH$\alpha$\,301)  &  $464\pm23$\phn  &  $9.18\pm0.45$  \\
22\,....  &  05 46 20.2  &  $-00$ 10 20  &  1.8  &  2MASS J05462018-0010197  &  $143\pm14$\phn  &  $2.28\pm0.22$  \\
23\,....  &  05 46 22.4  &  $-00$ 08 53  &  2.6  &  SSV\,68 (=LkH$\alpha$\,302)  & 
$295\pm18$\tablenotemark{f} \phn  &  $5.45\pm0.33$  \\ 24\,....  &  05 46 25.9  &  $-00$ 10
15  &  3.2  &  \nodata      &  $47\pm8$\phn\phn  &  $0.79\pm0.13$  \\ 25\,....  &  05 46
33.3  &  $+00$ 02 54  &  13.8 &  SSV\,13      &  $97\pm14$\phn  &  $2.02\pm0.29$     
\enddata
\scriptsize{
\tablenotetext{a}{Units of right ascension are hours, minutes, and seconds.
 Units of declination are degrees, arcminutes, and arcseconds.}
\tablenotetext{b}{Off-axis angle measured in arcminutes from the aim point on the ACIS--S3 chip, 
	$\alpha(2000.0)=$ 05$^{\rm h}$46$^{\rm m}$13\fs1, $\delta(2000.0)=$ --00\arcdeg09\arcmin57\farcs2.}
\tablenotetext{c}{Background-corrected source counts in the 0.5--10 keV energy band.  On-source counts were 
	measured through a circular aperture 5\arcsec\ in radius if $\theta \leq  7\arcmin$, and 10\arcsec\ in 
	radius if $\theta > 7\arcmin$. We note that source counts in the narrower 0.5--8.0 keV total band used by some authors, e.g., Feigelson et al. (2002), 
would be virtually identical to those given here for the wider 0.5--10.0 keV range.}  
\tablenotetext{d}{Corrected, using a monoenergetic (1.5 keV) exposure map, for the non-uniformity in camera sensitivity across the field of view.}
\tablenotetext{e}{Possible non-thermal background radio source, Source no. 1 of Bontemps et al. (1995).  See also Anglada et al. (1998).}
\tablenotetext{f}{Source located near the edge of a detector chip.}
}
\end{deluxetable}

\clearpage

\begin{deluxetable}{lcccccc}
\tablenum{2}
\tablewidth{0pc}
\tablecolumns{7}
\tabletypesize{\small}
\tablecaption{X$-$Ray Fluxes and Luminosities in L1630}
\tablehead{ 
\colhead{}        &  \colhead{Object}          & \colhead{R.A.\tablenotemark{a}}         &
\colhead{Decl.\tablenotemark{a}}         & \colhead{Count Rate\tablenotemark{b}}   &
\colhead{Flux at Earth\tablenotemark{c}} &  \colhead{$\log L_x$\tablenotemark{d}}        
\\
\colhead{Source}  & \colhead{Class}      & \colhead{(J2000)}                       &
\colhead{(J2000)}                        & \colhead{(counts ks$^{-1}$)}            &
\colhead{($10^{-14}$ \ecs)}              & \colhead{(\ergs)}
}
\startdata
SSV\,59\,...........  & I   &  05 46 04.8  &  $-00$ 14 17  &   $<0.20$  &        $<0.40$      & $<30.03$\phn\phn  \\
SSV\,60\,...........  & gK/M  &  05 46 08.4  &  $-00$ 14 25  & $<0.17$  &        $<0.34$      & $<29.96$\phn\phn  \\
SSV\,61\,...........  & wTTS  &  05 46 07.8  &  $-00$ 11 57  & $251.9\pm2.2$\phn\phn & $225\pm9$\phn\phn  & $31.86$    \\
SSV\,63W\,.......     & I   &  05 46 07.8  &  $-00$ 10 01  &   $2.65\pm0.24$  &  $5.30\pm0.52$      & $31.16$           \\
SSV\,63E\,.........   & I   &  05 46 08.5  &  $-00$ 10 03  &   $0.62\pm0.11$  &  $1.24\pm0.23$      & $30.52$           \\
SSV\,63NE\,......     & I/0 &  05 46 08.9  &  $-00$ 09 56  &   $0.39\pm0.10$  &  $0.78\pm0.20$      & $30.32$           \\
HH\,24-MMS\,...       & 0   &  05 46 08.3  &  $-00$ 10 43  &   $<0.17$  &        $<0.34$      & $<29.96$\phn\phn  \\
HH\,25-MMS\,...       & 0   &  05 46 07.3  &  $-00$ 13 30  &   $<0.22$  &        $<0.44$      & $<30.07$\phn\phn  \\
HH\,26-IR\,........   & I   &  05 46 03.9  &  $-00$ 14 53  &   $<0.22$  &        $<0.44$      & $<30.07$\phn\phn  \\
LBS\,17-H\,........   & 0   &  05 46 30.8  &  $-00$ 02 40  &   $<0.12$  &        $<0.24$      & $<29.81$\phn\phn 
\enddata
\scriptsize{
\tablenotetext{a}{Units of right ascension are hours, minutes, and seconds.
 Units of declination are degrees, arcminutes, and arcseconds.}
\tablenotetext{b}{Background-corrected source count rate, adjusted for non-uniform camera response. Upper limits 
	are for a detection cell of radius $r = 5\arcsec$ and are significant at the level of 2\,$\sigma$.}
\tablenotetext{c}{Flux in the 0.3--10 keV band.  Except for SSV\,61, we assume that ECF $= 2.00 \times 10^{-11}$ 
	\ecs\ per count s$^{-1}$.}
\tablenotetext{d}{Absorption-corrected X-ray luminosity or 2\,$\sigma$ upper limit in the 0.3--10 keV band, assuming
	a distance of $d = 450$ pc and, for all sources except SSV\,61, ECF $= 5.40 \times 10^{33}$ \ergs\ per count 
	s$^{-1}$. As noted in the text, SSV\,60 may be a normal, highly reddened late-K or early-M giant star 
	located behind L1630 at $d \approx 2$ kpc.  In general, the uncertainty in $L_x$ for each source is dominated 
	by the uncertainty in the distance of 15\% or more.} 
}
\end{deluxetable}

\end{document}